\documentclass[12pt,a4paper]{article}

\usepackage{amssymb}
\usepackage{amsmath}
\usepackage{epsfig}
\usepackage{graphicx}
\usepackage{enumerate}
\usepackage[footnotesize]{caption}
\newcommand{\mathsym}[1]{{}}

\newcommand{\ket}[1]{\left|#1\right\rangle}      
\newcommand{\bra}[1]{\left\langle #1\right|}     

\begin{document}

\newpage
\setcounter{page}{0}
\begin{titlepage}
\begin{flushright}
{\footnotesize AEI - 2010 - 018}
\end{flushright}
\vskip 1.5cm
\begin{center}
{\Large \textbf{Functional relations for the \\  six vertex model with domain wall boundary conditions}}\\
\vspace{1.5cm}
{\large W. Galleas} \\
\vspace{2cm}
{\it Max-Planck-Institut f\"ur Gravitationphysik \\
Albert-Einstein-Institut \\
Am M\"uhlenberg 1, 14476 Potsdam, Germany}\\
\end{center}
\vspace{1.5cm}

\begin{abstract}
In this work we demonstrate that the Yang-Baxter algebra can also be employed in order to derive
a functional relation for the partition function of the six vertex model with domain wall  boundary
conditions. The homogeneous limit is studied for small lattices and the properties determining the
partition function are also discussed.
\end{abstract}

\vspace{2.0cm}
\centerline{{\small PACS numbers:  05.50+q, 02.30.IK}}
\vspace{.1cm}
\centerline{{\small Keywords: Yang-Baxter equation, Functional equations, Domain wall boundaries}}
\vspace{2.0cm}
\centerline{{\small April 2010}}
\end{titlepage}

\tableofcontents

\section{Introduction}
The six vertex model with domain wall boundary conditions was introduced in \cite{kor} and since then
many connections with enumerative combinatorics and orthogonal polynomials theory have been unveiled \cite{vr0}. 
For instance, the problem of alternating sign matrices (ASM) and domino tilings enumeration is known
to have a close relationship with the six vertex model with this particular boundary condition \cite{vr2,vr2a,vr2b,vr2c}.
Moreover, the partition function of this model was also shown to correspond to a 
Schubert polynomial in \cite{vr3} and to a KP $\tau$ function in \cite{vr4}.

From the physical point of view the study of the partition function of this model reveals us
an interesting phenomena. Using the determinant representation found 
by Izergin \cite{vr5} and a Toda chain differential equation \cite{sogo} satisfied by this partition function,
it was shown in \cite{vr6} that the bulk free energy of the six vertex model with domain wall boundaries 
in the thermodynamical limit differs from the one with periodic boundary conditions. This fact 
has been also discussed in \cite{vr7} and raises the issue of the sensitivity of the six
vertex model in the thermodynamical limit with boundary conditions.

Domain wall boundary conditions were first introduced and studied within the scope of the
Quantum Inverse Scattering Method \cite{kor}, and they emerge naturally in the calculation
of correlations functions of quantum integrable systems \cite{qism2}. On the other hand, integrable systems
with an underlying Yang-Baxter symmetry have also been tackled by functional equations methods, which
are intimately connected with Baxter's commuting transfer matrix approach \cite{com}. However, since functional
methods usually do not provide the eigenvectors of the system, it is usually claimed that the calculation
of correlation functions is out of reach for functional equations methods. 

The use of the Yang-Baxter algebra in order to obtain functional equations was first introduced in
\cite{AF} and the aim of this paper is to show that the Yang-Baxter equation can still be explored in order to
obtain a functional relation for the partition function of the six vertex model
with domain wall boundaries.

This paper is organized as follows. In the section 2 we describe the six vertex model with domain wall
boundaries and its construction in terms of the Yang-Baxter algebra elements. In the section 3 we derive
a functional equation determining the partition function of the model and in section 4 and 5 we study its 
homogeneous limit and some particular solutions. Concluding remarks and open questions are discussed in the section 6 
and in the appendix A and B we present some extra results and technical details.

\section{The six vertex model with domain wall boundaries}

Vertex models in Statistical Mechanics were first introduced by L. Pauling aiming to describe the residual
entropy of ice \cite{lpauling}. These models are described in terms of a matrix $\mathcal{L}$ containing the
statistical weights  of the possible vertex configurations.

In a two dimensional rectangular lattice consisting of $M$ horizontal and $L$ vertical lines we have 
$L \times M$ intersection points and the intersection point of the
$i$-th horizontal and the $j$-th vertical line together with the four connecting edges is referred
as a vertex, as represented in the Fig. 1.

\begin{figure}[h]
\begin{center}
\includegraphics[width=16.2cm]{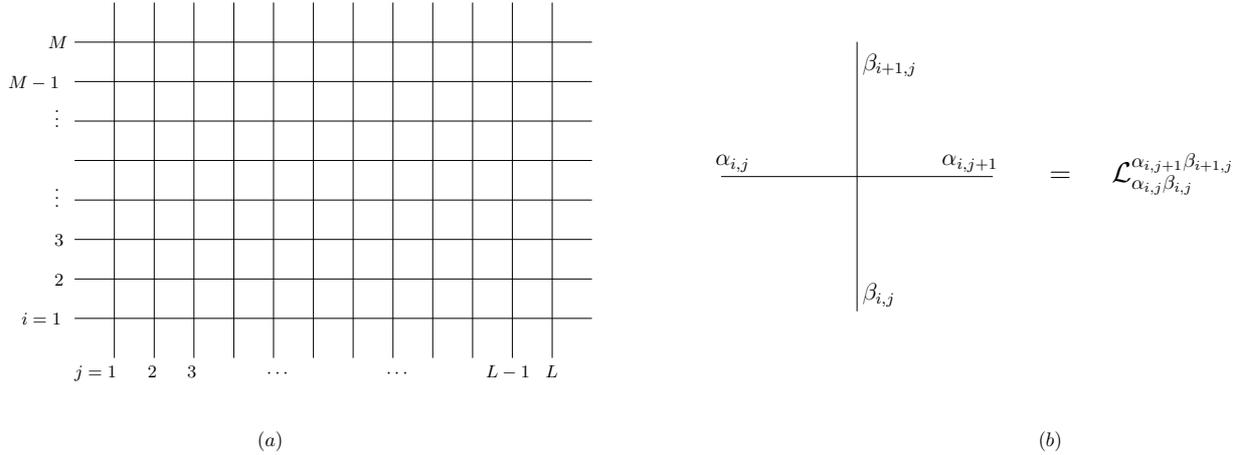}
\caption{(a) Rectangular lattice with $M$ lines and $L$ columns. (b) The vertex at position $(i,j)$ with its corresponding statistical weight.}
\end{center}
\end{figure}
The edge variables $\{ \alpha_{ij} , \alpha_{i j+1} , \beta_{i j} , \beta_{i+1 j} \}$ characterizes the configuration of the
vertex at position $(i,j)$ while $\mathcal{L}^{\alpha_{i,j+1} \beta_{i+1,j}}_{\alpha_{i,j} \beta_{i,j+1}}$ denotes its statistical weight.
The partition function of the model is thus obtained by summing the product of the statistical weights over all
possible configurations, i.e.
\begin{equation}
\label{zsum}
Z = \sum_{\{ \alpha , \beta \}} \prod_{i=1}^{M} \prod_{j=1}^{L} \mathcal{L}^{\alpha_{i,j+1} \beta_{i+1,j}}_{\alpha_{i,j} \beta_{i,j}}.
\end{equation}
For a more detailed discussion about vertex models and their applications in Statistical Mechanics see for instance \cite{bax} and references therein. 

In order to evaluate the sum (\ref{zsum}) it is usually assumed a particular boundary condition and the concept of domain wall
boundary conditions for the inhomogeneous six vertex model was introduced by Korepin in the Ref. \cite{kor}.
Turning our attention to the six vertex model, each edge variable $\alpha_{ij}$ and $\beta_{ij}$ can assume
two possible configurations under a certain restriction known as ice rule. These assumptions result in the following
$\mathcal{L}$-matrix 
\begin{equation}
\label{lm}
\mathcal{L} (\lambda)= \left( \begin{matrix}
a(\lambda) & 0 & 0 & 0 \\
0 & b(\lambda) & c(\lambda) & 0 \\
0 & c(\lambda) & b(\lambda) & 0 \\
0 & 0 & 0 & a(\lambda)
\end{matrix} \right)
\end{equation}
containing the statistical weights $a$, $b$ and $c$ of the allowed configurations. The weights are explicitly given by
$a(\lambda)=\sinh{(\lambda + \gamma)}$, $b(\lambda)=\sinh{(\lambda)}$ and $c(\lambda)=\sinh{(\gamma)}$ where $\gamma$ is the
anisotropy parameter and the complex variable $\lambda$ parametrizes the Yang-Baxter integrable manifold
\begin{equation}
\frac{a^2 + b^2 - c^2}{2 a b} = \Delta \;\;\;\;\;\;\;\;\;\;\;\;\;\;\; \Delta = \cosh{(\gamma)} .
\end{equation} 
Strictly speaking, the $\mathcal{L}$-matrix (\ref{lm}) satisfies the Yang-Baxter relation
\begin{equation}
\label{yb}
\mathcal{L}_{12}(\lambda - \mu) \mathcal{L}_{13}(\lambda - \nu) \mathcal{L}_{23}(\mu - \nu) =   
\mathcal{L}_{23}(\mu - \nu) \mathcal{L}_{13}(\lambda - \nu) \mathcal{L}_{12}(\lambda - \mu)
\end{equation}
where $\mathcal{L}_{ij} \in \mbox{End} \left( V_i \otimes V_j \right)$ and $V_i \cong \mathbb{C}^2$.
Consequently, the monodromy matrix 
\begin{equation}
\label{mono}
\mathcal{T}(\lambda, \{ \mu_{k} \}) = \mathcal{L}_{\mathcal{A} 1}(\lambda - \mu_{1}) \mathcal{L}_{\mathcal{A} 2}(\lambda - \mu_{2}) \dots \mathcal{L}_{\mathcal{A} L}(\lambda - \mu_{L}) 
\end{equation}
satisfies the relation
\begin{equation}
\label{rtt}
R(\lambda - \nu) \; \mathcal{T}(\lambda, \{ \mu_{k} \}) \otimes \mathcal{T}(\nu, \{ \mu_{k} \}) = 
\mathcal{T}(\nu, \{ \mu_{k} \}) \otimes \mathcal{T}(\lambda, \{ \mu_{k} \}) \; R(\lambda - \nu) 
\end{equation}
where $R(\lambda) = P \mathcal{L}(\lambda)$ and $P$ is the standard permutation matrix. 
The relation (\ref{rtt}) is
commonly referred as Yang-Baxter algebra and, together with the relation (\ref{yb}), they constitute the basis of the
Quantum Inverse Scattering Method \cite{qism1,qism2}.

Boundary conditions are an important ingredient in the formulation of lattice models and the case with domain wall boundary conditions
introduced in \cite{kor} requires that the boundary edges have a particular configuration respecting the six vertex model
symmetry.  For the six vertex model each edge variable $\{ \alpha_{ij} , \beta_{ij} \}$ can assume two possible configurations which
can be conveniently denoted by arrows pointing inwards or outwards, i.e.
\begin{eqnarray}
\alpha_{i,j} &=& \rightarrow \;\;\; \mbox{or} \;\;\; \leftarrow \nonumber \\
\beta_{i,j} &=& \downarrow \;\;\;\;\; \mbox{or} \;\;\; \uparrow
\end{eqnarray}
In this way domain wall boundary conditions consist of the restrictions
\begin{align}
\alpha_{i,1} &= \;\; \rightarrow&  \alpha_{i,L+1} =  &\leftarrow \nonumber \\
\beta_{1,j}  &= \;\; \downarrow&  \beta_{L+1,j}  =  &\uparrow 
\end{align}
where now we are considering a square lattice with $M=L$.
In order to make this situation more clear and intuitive, we have depicted a possible 
six vertex lattice configuration with domain wall boundaries in the Fig. 2.

\begin{figure}[h]
\begin{center}
\includegraphics[width=5cm]{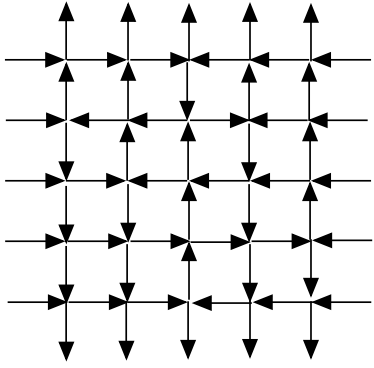}
\caption{A configuration with domain wall boundary conditions.}
\end{center}
\end{figure}
For the six vertex model the monodromy matrix (\ref{mono}) consist of a $2 \times 2$ matrix with operator valued entries.
Within the framework of the Quantum Inverse Scattering Method it can be conveniently denoted as
\begin{equation}
\label{abcd}
\mathcal{T}(\lambda , \{ \mu_{k} \} ) = \left( \begin{matrix}
A(\lambda , \{ \mu_{k} \} ) & B(\lambda , \{ \mu_{k} \} ) \\
C(\lambda , \{ \mu_{k} \} ) & D(\lambda , \{ \mu_{k} \} )
\end{matrix} \right)
\end{equation}
and it was shown in \cite{kor} that the partition function (\ref{zsum}) of the inhomogeneous six vertex model with domain wall boundary
conditions can be written as
\begin{equation}
\label{pf}
Z_{DW} = \bra{\bar{0}} \prod_{j=1}^L B(\lambda_{j}, \{ \mu_k \} ) \ket{0}
\end{equation}
or equivalently 
\begin{equation}
Z_{DW} = \bra{0} \prod_{j=1}^L C(\lambda_{j}, \{ \mu_k \} ) \ket{\bar{0}} \; ,
\end{equation}
where $\ket{0}$ and $\ket{\bar{0}}$ denote the usual ferromagnetic states
\begin{eqnarray}
\label{vac}
\ket{0} = \bigotimes_{i=1}^{L} 
\left( \begin{matrix}
1 \\
0 
\end{matrix} \right)
\;\;\; \mbox{and} \;\;\;
\ket{\bar{0}} = \bigotimes_{i=1}^{L} 
\left( \begin{matrix}
0 \\
1 
\end{matrix} \right).
\end{eqnarray}
In the next sections we shall demonstrate how the Yang-Baxter algebra can be employed in order to obtain  
a functional relation for the partition function (\ref{pf}).

\section{Yang-Baxter algebra and functional relations}

The Yang-Baxter algebra is a corner stone of the Quantum Inverse Scattering Method and it has been successfully explored in order
to construct integrable systems and to extract exact results from them. More recently, it was shown in \cite{AF} that the Yang-Baxter
algebra can also render functional relations determining the spectrum of spin chains with non-diagonal twisted and open boundary
conditions.

In the Ref. \cite{kor} Korepin obtained a recursion relation for the partition function (\ref{pf}) whose solution was later on 
given by Izergin \cite{izer} in terms of a determinant formula. Our purpose here is to demonstrate that the Yang-Baxter
algebra also provide us a functional equation determining the partition function of the six vertex model with domain
wall boundaries. 
In order to do so we first need to recall some properties exhibited by the six vertex model monodromy matrix (\ref{mono}).

Let $\ket{0}_j$ and $\ket{\bar{0}}_j$ be the states $\left( \begin{matrix} 1 \\ 0 \end{matrix} \right)$ and
$\left( \begin{matrix} 0 \\ 1 \end{matrix} \right)$ respectively, acting on the $j$-th space of the tensor product
$V_1 \otimes V_2 \otimes \dots \otimes V_j \otimes \dots \otimes V_L$. The action of $\mathcal{L}_{\mathcal{A} j}$ on these states 
yields us a triangular matrix. For instance we have
\begin{equation}
\label{t1}
\mathcal{L}_{\mathcal{A} j}(\lambda) \ket{0}_j = \left( \begin{matrix}
a(\lambda) \ket{0}_j & \dagger \\
0 & b(\lambda) \ket{0}_j
\end{matrix} \right)
\end{equation}
and analogously
\begin{equation}
\label{t2}
\mathcal{L}_{\mathcal{A} j}(\lambda) \ket{\bar{0}}_j = \left( \begin{matrix}
b(\lambda) \ket{\bar{0}}_j & 0 \\
\dagger & a(\lambda) \ket{\bar{0}}_j
\end{matrix} \right) \; ,
\end{equation}
where the symbol $\dagger$ stands for a non-null value.

Due to its definition, the monodromy matrix (\ref{mono}) inherits the triangular properties (\ref{t1}) and (\ref{t2}). In this way the monodromy
matrix elements satisfy the relations
\begin{align}
\label{action}
A(\lambda, \{ \mu_k \} ) \ket{0} &= \prod_{j=1}^L a(\lambda - \mu_j) \ket{0}&   B(\lambda, \{ \mu_k \} ) \ket{0} = & \dagger \nonumber \\
C(\lambda, \{ \mu_k \} ) \ket{0} &= 0&    D(\lambda, \{ \mu_k \} ) \ket{0} = & \prod_{j=1}^L b(\lambda - \mu_j) \ket{0} \; ,
\end{align}
and
\begin{align}
\label{actionb}
A(\lambda, \{ \mu_k \} ) \ket{\bar{0}} &= \prod_{j=1}^L b(\lambda - \mu_j) \ket{0}&   B(\lambda, \{ \mu_k \} ) \ket{\bar{0}} = & 0 \nonumber \\
C(\lambda, \{ \mu_k \} ) \ket{\bar{0}} &= \dagger&    D(\lambda, \{ \mu_k \} ) \ket{\bar{0}} = & \prod_{j=1}^L a(\lambda - \mu_j) \ket{\bar{0}} \; .
\end{align}

Another fundamental ingredient in the method considered here is the Yang-Baxter algebra. The relation (\ref{rtt}) encodes several commutation
rules for the elements of the monodromy matrix and among them we shall
make use of the following ones
\begin{eqnarray}
\label{abc}
A(\lambda , \{ \mu_k \}) B(\nu , \{ \mu_k \}) &=& \frac{a(\nu - \lambda)}{b(\nu - \lambda )} B(\nu , \{ \mu_k \}) A(\lambda , \{ \mu_k \}) - 
\frac{c(\nu - \lambda)}{b(\nu - \lambda )} B(\lambda , \{ \mu_k \}) A(\nu , \{ \mu_k \}) \nonumber \\
D(\lambda , \{ \mu_k \}) B(\nu , \{ \mu_k \}) &=& \frac{a(\lambda - \nu)}{b(\lambda - \nu)} B(\nu , \{ \mu_k \}) D(\lambda , \{ \mu_k \}) - 
\frac{c(\lambda - \nu)}{b(\lambda - \nu )} B(\lambda , \{ \mu_k \}) D(\nu , \{ \mu_k \}) \nonumber \\ 
\left[ C(\lambda , \{ \mu_k \}) , B(\nu , \{ \mu_k \}) \right] &=& \frac{c(\lambda - \nu)}{b(\lambda - \nu)} \left[ A(\nu , \{ \mu_k \}) D(\lambda , \{ \mu_k \}) - A(\lambda , \{ \mu_k \}) D(\nu , \{ \mu_k \}) \right] \nonumber \\
B(\lambda , \{ \mu_k \})  B(\nu , \{ \mu_k \}) &=&  B(\nu , \{ \mu_k \}) B(\lambda , \{ \mu_k \})   \; .
\end{eqnarray}

In the Refs. \cite{qism1} and \cite{kor} it was shown that the operator $B(\lambda , \{ \mu_k \})$ plays the role of raising operator with respect
to the pseudo-vacuum state $\ket{0}$ while the operator $C(\lambda , \{ \mu_k \})$ acts as a lowering operator. Furthermore, it was also demonstrated in 
\cite{kor} that
\begin{eqnarray}
\label{cbb}
C(\lambda_0 , \{ \mu_k \}) \prod_{i=1}^{n}  B(\lambda_i , \{ \mu_k \}) \ket{0} &=& \sum_{i=1}^{n} M_i \prod_{\stackrel{j=1}{j \neq i}}^{n}  B(\lambda_j , \{ \mu_k \}) \ket{0} \nonumber \\ 
&+& \sum_{1 \leq i < j \leq n} N_{ji} B(\lambda_0 , \{ \mu_k \}) \prod_{\stackrel{l=1}{l \neq i,j}}^{n}  B(\lambda_l , \{ \mu_k \}) \ket{0} 
\end{eqnarray}
with
\begin{eqnarray}
\label{mn}
M_i &=& \frac{c(\lambda_i - \lambda_0)}{b(\lambda_i - \lambda_0)} \prod_{l=1}^{L} a(\lambda_0 - \mu_l) b(\lambda_i - \mu_l) \prod_{\stackrel{k=1}{k \neq i}}^{n} \frac{a(\lambda_i - \lambda_k)}{b(\lambda_i - \lambda_k)} \frac{a(\lambda_k - \lambda_0)}{b(\lambda_k - \lambda_0)} \nonumber \\
&+& \frac{c(\lambda_0 - \lambda_i)}{b(\lambda_0 - \lambda_i)} \prod_{l=1}^{L} a(\lambda_i - \mu_l) b(\lambda_0 - \mu_l) \prod_{\stackrel{k=1}{k \neq i}}^{n} \frac{a(\lambda_0 - \lambda_k)}{b(\lambda_0 - \lambda_k)} \frac{a(\lambda_k - \lambda_i)}{b(\lambda_k - \lambda_i)}
\end{eqnarray} 
\begin{eqnarray}
\label{mn1}
N_{ji} &=& \frac{c(\lambda_0 - \lambda_j)}{b(\lambda_0 - \lambda_j)} \frac{c(\lambda_i - \lambda_0)}{b(\lambda_i - \lambda_0)} \frac{a(\lambda_j - \lambda_i)}{b(\lambda_j - \lambda_i)} \prod_{l=1}^{L} a(\lambda_i - \mu_l) b(\lambda_j - \mu_l) \prod_{\stackrel{m=1}{m \neq i,j}}^{n} \frac{a(\lambda_j - \lambda_m)}{b(\lambda_j - \lambda_m)} \frac{a(\lambda_m - \lambda_i)}{b(\lambda_m - \lambda_i)} \nonumber \\
&+& \frac{c(\lambda_0 - \lambda_i)}{b(\lambda_0 - \lambda_i)} \frac{c(\lambda_j - \lambda_0)}{b(\lambda_j - \lambda_0)} \frac{a(\lambda_i - \lambda_j)}{b(\lambda_i - \lambda_j)} \prod_{l=1}^{L} a(\lambda_j - \mu_l) b(\lambda_i - \mu_l) \prod_{\stackrel{m=1}{m \neq i,j}}^{n} \frac{a(\lambda_i - \lambda_m)}{b(\lambda_i - \lambda_m)} \frac{a(\lambda_m - \lambda_j)}{b(\lambda_m - \lambda_j)} \nonumber \\
\end{eqnarray}
for any number $n$ of operators $B(\lambda_i , \{ \mu_k \})$. The demonstration of (\ref{cbb})-(\ref{mn1}) only makes use of the commutation rules (\ref{abc})
together with the relations (\ref{action}).

At this stage we have gathered most of the ingredients required to obtain a functional relation for the partition function (\ref{pf}). In order to proceed we 
look to the relation (\ref{cbb}) with $n=L+1$ and act with the dual vector $\bra{\bar{0}}$ on its left hand side. By doing so we obtain the
following relation
\begin{eqnarray}
\label{cblz}
\bra{\bar{0}} C(\lambda_0 , \{ \mu_k \}) && \prod_{i=1}^{L+1}  B(\lambda_i , \{ \mu_k \}) \ket{0} = 
\sum_{i=1}^{L+1} M_i \; Z(\lambda_1, \dots, \lambda_{i-1}, \lambda_{i+1}, \dots, \lambda_{L+1} ) \nonumber \\
&+& \sum_{1 \leq i < j \leq L+1} N_{ji} \; Z(\lambda_0 , \lambda_1, \dots, \lambda_{i-1}, \lambda_{i+1}, \dots, \lambda_{j-1}, \lambda_{j+1}, \dots, \lambda_{L+1} ) \nonumber \\
\end{eqnarray}
where $Z(\lambda_1, \dots, \lambda_{L})$ denotes the partition function (\ref{pf}) omitting the dependence with the variables $\{ \mu_k \}$.

We shall refer to $\displaystyle \prod_{i=1}^{n}  B(\lambda_i , \{ \mu_k \}) \ket{0}$ as Bethe vectors and it was also shown in \cite{kor} that
\begin{equation}
\prod_{j=1}^{L}  B(\lambda_j , \{ \mu_k \}) \ket{0} = Z(\lambda_1 , \dots , \lambda_L ) \ket{\bar{0}}
\end{equation}
which implies in 
\begin{equation}
\label{z0}
\prod_{j=1}^{L+1}  B(\lambda_j , \{ \mu_k \}) \ket{0} = 0
\end{equation}
due to the relations (\ref{actionb}). The Bethe vectors enjoy the property of being highest weight $SU(2)$ vectors \cite{hw,kar} and a careful examination of (\ref{cblz}), taking
into account the relation (\ref{z0}), reveals that its left hand side vanishes. In other words, the consistency of the Yang-Baxter algebra
with the highest weight property  of the Bethe vectors results in the following functional equation for (\ref{pf}),
\begin{eqnarray}
\label{FZ}
&&\sum_{i=1}^{L+1} M_i \; Z(\lambda_1, \dots, \lambda_{i-1}, \lambda_{i+1}, \dots, \lambda_{L+1} ) \nonumber \\
&+& \sum_{1 \leq i < j \leq L+1} N_{ji} \; Z(\lambda_0 , \lambda_1, \dots, \lambda_{i-1}, \lambda_{i+1}, \dots, \lambda_{j-1}, \lambda_{j+1}, \dots, \lambda_{L+1} ) =0 \; ,
\end{eqnarray}
where the functions $M_i$ and $N_{ji}$ are given by the relations (\ref{mn}) and (\ref{mn1}) with $n=L+1$. We have verified the validity of the relation (\ref{FZ}) by explicit computing the partition function  $Z(\lambda_1, \dots,\lambda_{L})$ for $L=1,2,3,4,5$ using the definition (\ref{pf}), and in what follows we shall
discuss the properties of the Eq. (\ref{FZ}) as well as some particular solutions.

\section{The homogeneous limit}
The results of the previous sections focus on an inhomogeneous lattice whose statistical weights are parametrized by variables 
$\{ \lambda_k \}$ and $\{ \mu_k \}$. In this section we analyze explicitly the homogeneous limit
\begin{equation}
\lambda_k \rightarrow \lambda \;\;\;\;\;\;\;\;\;\;\;\; \mu_k \rightarrow \mu
\end{equation}
in the Eq. (\ref{FZ}) for $L=1,2$ which already unveils the properties that uniquely determine the partition function (\ref{pf}).

The case $L=1$ is trivial since the partition function $Z(\lambda)$ can be promptly written down from the definitions
(\ref{lm}), (\ref{mono}) and (\ref{pf}). However it is worthwhile to look at (\ref{FZ}) with $L=1$ for illustrative purposes.
In that case we have the equation
\begin{equation}
\label{L1}
M_1 Z(\lambda_2) + M_2 Z(\lambda_1) + N_{21} Z(\lambda_0) = 0 \; ,
\end{equation}
and by inspecting the relations (\ref{mn}) and (\ref{mn1}) we find in general that the limit $\mu_k \rightarrow \mu$
can be easily obtained. By way of contrast the limit $\lambda_k \rightarrow \lambda$ is highly non-trivial due to the presence of
poles in the functions $M_i$ and $N_{ji}$ when two variables $\lambda_k$ coincide. In spite of that, this limit can be taken using 
L'Hopital's rule and we shall see that the polynomial structure and the asymptotic behavior discussed in the appendix A
also plays an important role in the unique determination of the the partition function $Z(\lambda_1, \dots,\lambda_{L})$.

Considering the limits $\lambda_0 , \lambda_1 , \lambda_2 \rightarrow \lambda$ and $\mu_1 \rightarrow \mu$ 
in the Eq. (\ref{L1}) we obtain the following 
differential equation,
\begin{eqnarray}
\label{ed1}
\left[ 1 - \frac{2 q x}{(q + q^{-1})} \right] \frac{d Z}{dx} + \frac{x}{2}\left[ 1 - \frac{4 q x}{(q+q^{-1})} + q^2 x^2 \right] \frac{d^2 Z}{dx^2} = 0
\end{eqnarray}
expressed in terms of the variables $x=e^{2(\lambda - \mu)}$ and $q=e^{\gamma}$. The Eq. (\ref{ed1}) possess the general solution
\begin{equation}
Z(x) = K_1 + K_2 \left[ (1+q^2)(xq^2 - x^{-1}) - 4 q^2 \log{x} \right] \; ,
\end{equation}
where $K_1$ and $K_2$ are arbitrary constants. However, the polynomial structure discussed in the appendix A requires that $K_2 = 0$ and we are thus left with
$Z(x) = K_1$. The constant $K_1$ is then fixed by the asymptotic behavior, see appendix A, and for $L=1$ we end up with
\begin{equation}
Z(x) = \frac{q - q^{-1}}{2}.
\end{equation}
As mentioned before, the case $L=1$ is a trivial case and we now turn our attention to the case $L=2$. In that case the Eq. (\ref{FZ}) reads
\begin{eqnarray}
\label{L2}
&& M_1 \; Z(\lambda_2 , \lambda_3) + M_2 \; Z(\lambda_1 , \lambda_3) + M_3 \; Z(\lambda_1 , \lambda_2) \nonumber \\
&&+ \; N_{21} \; Z(\lambda_0 , \lambda_3) + N_{31} \; Z(\lambda_0 , \lambda_2) + N_{32} \; Z(\lambda_0 , \lambda_1) =0
\end{eqnarray}
and again we can use L'Hopital's rule in order to evaluate the limit $\lambda_0 , \lambda_1 , \lambda_2 , \lambda_3 \rightarrow \lambda$ and
$\mu_1 , \mu_2 \rightarrow \mu$. In terms of the variables $x_{i} = e^{2(\lambda_i - \mu_i)}$, we have that $Z(x_1 ,x_2)=\frac{\bar{Z}(x_1 , x_2)}{\sqrt{x_1 x_2}}$
where the function $\bar{Z}(x_1 ,x_2)$ is a polynomial
of degree $1$ in each variable separately as discussed in the appendix A. Thus, in the homogeneous limit, the function $\bar{Z}$ is a polynomial of order $2$ in the variable $x$.
This fact implies in $\displaystyle \frac{d^n \bar{Z}}{dx^n}=0$ for $n > 2$ and taking this into account, when considering the 
homogeneous limit in (\ref{L2}), we obtain the equation
\begin{equation}
\label{ed2}
\Phi_0 (x) \bar{Z}(x) + \Phi_1 (x) \frac{d \bar{Z}(x)}{dx} + \Phi_2 (x) \frac{d^2 \bar{Z}(x)}{dx^2} = 0  
\end{equation}
where
\begin{eqnarray}
\Phi_0 (x) &=& -4 q^2 (1 + q^2 + q^4 ) + 6 q^4 (1+q^2) x + 12 q^6 x^2 -6 q^6 (1+q^2) x^3 \nonumber \\
\Phi_1 (x) &=& -( 1 + 2 q^2 + 2 q^4 + q^6 ) + 4 q^2 (1 + q^2 +q^4)x -12 q^6 x^3 + q^4 (-1 + 4 q^2 + 4 q^4 -q^6) x^4 \nonumber \\
\Phi_2 (x) &=& (1 - q^2 - q^4 + q^6) x - 2 q^2 (1 - 2 q^2 + q^4 )(1 + q^2 x^2 )x^2 + q^4 (1 - q^2 - q^4 + q^6) x^5 \; . \nonumber \\
\end{eqnarray}

Looking to the polynomial solution of (\ref{ed2}), we can generically write 
\begin{equation}
\label{anz}
\bar{Z}(x) = k_2 x^2 + k_1 x + k_0
\end{equation}
where $k_0$ , $k_1$ and $k_2$ are arbitrary constants.  In this way by replacing (\ref{anz}) into (\ref{ed2})
we find 
\begin{eqnarray}
k_0 = \frac{k_2}{q^2}  \;\;\;\;\;\;\;\;\; \mbox{and}  \;\;\;\;\;\;\;\; k_1 = - \frac{4 k_2}{(1+q^2)} \; ,
\end{eqnarray}
and the constant $k_2$ is then fixed by the asymptotic behavior of $\bar{Z}$ which can be found in the appendix A. It turns out that
\begin{equation}
k_2 = \frac{1}{16} (q - q^{-1})^{2} (1 + q^{2}) \; ,
\end{equation}
and we close this section summarizing our results. 

Although we have considered only solutions in the homogeneous limit, this analysis suggests
that the partition function $Z(\lambda_1 , \dots , \lambda_L)$ is determined by the following properties:
\begin{enumerate}[(i)]
\item Functional relation (\ref{FZ});
\item Polynomial structure: 
\begin{equation}
\displaystyle Z(\lambda_1 , \dots , \lambda_L) = \frac{\bar{Z}(x_1 , \dots , x_L)}{\displaystyle \prod_{i=1}^{L} x_{i}^{\frac{L-1}{2}}}
\end{equation}
where $\bar{Z}(x_1 , \dots , x_L)$ is a polynomial of degree $L-1$ in each variable $x_i$ separately;
\item Asymptotic behavior $\bar{Z}(x_1 , \dots , x_L) \sim \frac{(q-q^{-1})^L}{2^{L^2}} [ L ]_{q^2} ! \; (x_{1} \dots x_{L})^{L-1} $ as $x_i \rightarrow \infty$.
\end{enumerate}
Here $[ L ]_{q^2} !$ corresponds to the $q$-factorial function whose definition is given in the appendix A, and in the next section we shall discuss
the validity of the properties (i)-(iii) for the non-homogeneous case.

\section{Non-homogeneous solutions}
In order to analyze the role of the conditions (i)-(iii) given in the previous section we shall now consider the case with $\mu_{k} = 0$, 
while $\lambda_{k}$ are kept arbitrary, in addition to the homogeneous limit discussed previously. Let us start again
looking to the case $L=1$ where the partition function $Z$ is constrained by the Eq. (\ref{L1}). In that case the polynomial structure (ii) implies that $Z$ is a constant
and it solves the Eq. (\ref{L1}) due to the identity                    
\begin{equation}
M_1 + M_2 + N_{21} = 0 \; ,
\end{equation}
which holds even when $\mu_{1}$ is kept arbitrary. That constant is then determined by the property (iii) similarly to the homogeneous case
discussed in the previous section.

For the case $L=2$, the partition function $Z(\lambda_1 , \lambda_2)$ satisfies the Eq. (\ref{L2}) and the polynomial structure 
(ii) implies in the following general form for the partition function,
\begin{equation}
\label{pol12}
Z(\lambda_1 , \lambda_2) = \sum_{m,n = -1}^{1} h_{m,n} \; e^{m \lambda_1 + n \lambda_2}
\end{equation}
where $h_{m,n}$ are arbitrary coefficients.
By replacing the expression (\ref{pol12}) in the Eq. (\ref{L2}) and equating the coefficients we find constraints for $h_{m,n}$
which are solved by
\begin{eqnarray}
\label{coef12}
h_{-1,-1} &=& \frac{h_{1,1}}{q^2} \;\;\;\;\;\;\;\;\;\;\;\;\;\;\;\;\;\;\;\;\;\;\; h_{0,0}= 0 \nonumber \\
h_{-1,0} &=& 0 \;\;\;\;\;\;\;\;\;\;\;\;\;\;\;\;\;\;\;\;\;\;\;\;\;\;\;\; h_{0,1}= 0 \nonumber \\
h_{-1,1} &=& -\frac{2}{(1+q^2)} h_{1,1} \;\;\;\;\;\;\; h_{1,-1}= -\frac{2}{(1+q^2)} h_{1,1} \nonumber \\
h_{0,-1} &=& 0 \;\;\;\;\;\;\;\;\;\;\;\;\;\;\;\;\;\;\;\;\;\;\;\;\;\;\;\; h_{1,0}= 0 \; ,
\end{eqnarray}
while the leading term coefficient $h_{1,1}$ is then determined by the condition (iii),
\begin{equation}
h_{1,1} = \frac{1}{16} (q - q^{-1})^{2} (1 + q^{2}) .
\end{equation}

The fact that the Eq. (\ref{FZ}) determines all coefficients of the polynomial $Z(\lambda_1 , \dots, \lambda_L)$, except for one,
would be expected since the Eq. (\ref{FZ}) is invariant under the transformation $Z \rightarrow \Omega \; Z$ where $\Omega$ is an arbitrary
constant. Thus the role of the condition (iii) is the determination of the leading term coefficient of the polynomial $Z(\lambda_1 , \dots, \lambda_L)$.
The partition function $Z(\lambda_1 , \lambda_2)$ is also expected to be symmetric in the variables $\lambda_1$ and $\lambda_2$, i.e. 
$Z(\lambda_1 , \lambda_2) = Z(\lambda_2 , \lambda_1)$, due to the definition (\ref{pf}) and the commutation relations (\ref{abc}). Though we have
not made this assumption, the solution of the Eq. (\ref{L2}) given by (\ref{pol12}) and (\ref{coef12}) is manifested a symmetric polynomial reflecting
that this property is already incorporated in the Eq. (\ref{FZ}).

Turning our attention to the case $L=3$ the Eq. (\ref{FZ}) reads
\begin{eqnarray}
\label{L3}
M_1 Z(\lambda_2 , \lambda_3 , \lambda_4 ) &+& M_2 Z(\lambda_1 , \lambda_3 , \lambda_4 ) + M_3 Z(\lambda_1 , \lambda_2 , \lambda_4 ) + M_4 Z(\lambda_1 , \lambda_2 , \lambda_3 ) \nonumber \\
&+& N_{21} Z(\lambda_0 , \lambda_3 , \lambda_4 ) + N_{31} Z(\lambda_0 , \lambda_2 , \lambda_4 ) + N_{41} Z(\lambda_0 , \lambda_2 , \lambda_3 ) \nonumber \\
&+& N_{32} Z(\lambda_0 , \lambda_1 , \lambda_4 ) + N_{42} Z(\lambda_0 , \lambda_1 , \lambda_3 ) + N_{43} Z(\lambda_0 , \lambda_1 , \lambda_2 ) = 0
\end{eqnarray}
where the coefficients $M_i$ and $N_{ji}$ are given by the Eqs. (\ref{mn}) and (\ref{mn1}). Restricting ourselves to the polynomial solution of (\ref{L3}) characterized by the property (ii), the partition function $Z(\lambda_1 , \lambda_2 , \lambda_3)$ can be generically written as
\begin{equation}
\label{pol123}
Z(\lambda_1 , \lambda_2 , \lambda_3) = \sum_{m,n,o = -2}^{2} h_{m,n,o} \; e^{m \lambda_1 + n \lambda_2 + o \lambda_3 } \; ,
\end{equation}
which can be directly substituted in the Eq. (\ref{L3}). By doing so we find constraints fixing all the coefficients $h_{m,n,o}$ 
except for one, i.e. the leading term coefficient $h_{2,2,2}$, which is then fixed by the property (iii)
\begin{equation}
h_{2,2,2} = \frac{1}{2^9} (q - q^{-1})^3 (1 + q^2) (1 + q^2 + q^4) \; .
\end{equation}
In order to avoid an overcrowded section we have collected in the appendix B the non-null coefficients $h_{m,n,o}$
contained in the expression (\ref{pol123}). As we also observed for the
case $L=2$, the resulting expression (\ref{pol123}) is symmetric under the exchange of variables as a direct consequence of the Eq. (\ref{FZ}). 

We close this section remarking that although we have not rigorously proved that the conditions (i)-(iii) uniquely defines the partition function (\ref{pf}), the
explicit computations performed in this section and in the previous one, using solely the properties (i)-(iii), strongly suggests that those properties completely
determines the partition function $Z(\lambda_1 , \dots , \lambda_L)$.

\section{Concluding remarks}

In this paper we have obtained a functional equation for the partition function of the six vertex
model with domain wall boundary condition. The main ingredient in our derivation is the 
Yang-Baxter algebra pointing out a novel branch of exploration of this algebra.

Although we have explicitly verified the validity of the functional equation (\ref{FZ}) for $L=1,2,3,4,5$ , it would
be still interesting to provide a proof of the Izergin determinant representation \cite{vr5}. It is also 
worthwhile to remark that functional equations for this partition function have already been presented
in the literature \cite{strog} when the anisotropy parameter $q$ is a root of unity. The Eq. (\ref{FZ}) is valid
for general values of the anisotropy parameter $q$ and  certainly it would be interesting to investigate the connections
between our equation and the ones of \cite{strog}.

Different equations and representations for this partition function have also being investigated in the literature \cite{vr3,sogo,strog,pronko}
usually obtained from Izergin determinant formula. Our analysis do not rely on the determinant formula and  we hope the
Eq. (\ref{FZ}) still allows one to investigate alternative representations.

A careful examination of Eq. (\ref{FZ}), or the particular case depicted in (\ref{L2}), suggests that we have a relation between the
partition function in the complete homogeneous limit and the one in the partial homogeneous limit. This partial homogeneous
limit has been discussed in \cite{zeil} associated to the proof of the refined alternating sign matrix conjecture, and in \cite{pronko1}
related to the six vertex model artic curve. The examination of (\ref{FZ}) then suggests that such an equation relating partial and 
complete homogeneous limit  could be obtained by considering $\lambda_1 , \lambda_2 , \dots , \lambda_{L+1} \rightarrow \lambda$
while keeping $\lambda_0$ fixed.

Within the framework of the Quantum Inverse Scattering Method, the study of partition functions with domain wall boundary conditions
shares many aspects with the computation of scalar products and correlation functions \cite{qism2}. Given this similarity we hope
to report on functional equations for scalar products and correlation functions in a future publication.

\newpage
\addcontentsline{toc}{section}{Appendix A}
\section*{\bf Appendix A: Polynomial structure and asymptotic behavior.}
\setcounter{equation}{0}
\renewcommand{\theequation}{A.\arabic{equation}}
As it was shown in \cite{kor} and \cite{AF}, the definitions (\ref{lm}), (\ref{mono}) and (\ref{abcd}) allow us to study the dependence
of the operator $B(\lambda,\{ \mu_k \})$ with the spectral parameters. In terms of the variables $x_i = e^{2(\lambda_i - \mu_i)}$, it turns
out that
\begin{equation}
\label{ex}
B(\lambda_i , \{ \mu_k \}) = \frac{1}{x_{i}^{\frac{L-1}{2}}} \left[  f_{L-1}^{(i)} x_{i}^{L-1} + f_{L-2}^{(i)} x_{i}^{L-2} + \dots + f_{1}^{(i)} x_{i} + f_0^{(i)}  \right]
\end{equation}
with $L$ operator coefficients $f_{\alpha}$. The leading term $f_{L-1}$ can be written down explicitly due to the structure of (\ref{lm}) and (\ref{mono}), and it is 
given by 
\begin{equation}
f_{L-1}^{(i)} = 2^{-L} q^{\frac{L-3}{2}} (q^2-1) \exp{\left[(L-1) \mu_{i} - \sum_{j=1}^{L} \mu_{j} \right]} \sum_{j=1}^{L} e^{\mu_j} P_j
\end{equation}
where the operators $P_j$ are defined as
\begin{equation}
\label{pj}
P_j = \bigotimes_{l=1}^{j-1} K \otimes X^{-} \otimes \bigotimes_{l=j+1}^{L} K^{-1} \; .
\end{equation}
Together with $\displaystyle X^{+} = \left(\begin{matrix} 0 & 1 \\ 0 & 0 \end{matrix} \right)$ 
the matrices
\begin{equation}
\label{a4}
X^{-} = \left(\begin{matrix}
0 & 0 \\
1 & 0 \end{matrix} \right) \;\;\;\;\;\; \mbox{and} \;\;\;\;\;\;
K = \left(\begin{matrix}
q^{\frac{1}{2}} & 0 \\
0 & q^{-\frac{1}{2}} \end{matrix} \right) 
\end{equation}
generate the $q$-deformed $su(2)$ algebra,
\begin{eqnarray}
\label{qsu2}
K X^{\pm} K^{-1} &=& q^{\pm} X^{\pm} \nonumber \\
\left[ X^{+} , X^{-} \right] &=& \frac{K^2 - K^{-2}}{q-q^{-1}} \; .
\end{eqnarray}

Looking to the definition (\ref{pf}) we can readly see that the dependence of
$Z(\lambda_1 , \dots , \lambda_{L})$ with the spectral parameters follows directly from the dependence of
$\displaystyle \prod_{i=1}^{L} B(\lambda_i , \{ \mu_k \})$ with $\{ \lambda_j \}$ and $\{ \mu_k \}$ since the vectors $\ket{0}$
and $\ket{\bar{0}}$ do not depend on them.  Thus considering the definition (\ref{pf}) and (\ref{ex}) we can write
\begin{equation}
\label{a6}
\displaystyle Z(\lambda_1 , \dots , \lambda_L) = \frac{\bar{Z}(x_1 , \dots , x_L)}{\displaystyle \prod_{i=1}^{L} x_{i}^{\frac{L-1}{2}}}
\end{equation}
where $\bar{Z}(x_1 , \dots , x_L)$ consists of a polynomial of degree $L-1$ in each variable $x_i$ independently.

In their turn the coefficients $f_{L-1}^{(i)}$ govern the asymptotic behavior of $\bar{Z}$. In the limit $x_i \rightarrow \infty$ only the leading term
coefficients contribute and from (\ref{ex}) and (\ref{a6}) we obtain 
\begin{equation}
\label{a7}
\bar{Z}(x_1 , \dots , x_L) \sim \bra{\bar{0}} \prod_{i=1}^{L} f_{L-1}^{(i)} \ket{0}  \;\;\; (x_{1} \dots x_{L})^{L-1} \;\;\;\;\;\;\;\;\; \mbox{as} \;\; x_i \rightarrow \infty \; .
\end{equation}
The product $\displaystyle \prod_{i=1}^{L} f_{L-1}^{(i)}$ can be worked out using the following properties exhibited by the operators $P_j$,
\begin{eqnarray}
\label{pp1}
P_i P_j &=& q^2 P_j P_i \;\;\;\;\;\;\;\;\;\; (i < j) \\
\label{pp2}
P_i^2 &=& 0 \; ,
\end{eqnarray}
which can be derived with the help of the relations (\ref{pj})-(\ref{qsu2}). 

Considering the property (\ref{pp2}), we then have
\begin{equation}
\label{ff1}
\prod_{i=1}^{L} f_{L-1}^{(i)} = 2^{-L^2} q^{\frac{L(L-3)}{2}} (q^2 -1)^L \sum_{\{ a_j \}} P_{a_{1}} P_{a_{2}} \dots P_{a_{L}}
\end{equation}
where $\displaystyle \sum_{\{ a_j \}}$ denotes a summation over $\{ a_j \}$ with each index $a_j$ ranging from $1$ to $L$ under the constraint
$a_{1} \neq a_{2} \neq \dots \neq a_L$. The terms of $\displaystyle \sum_{\{ a_j \}} P_{a_{1}} P_{a_{2}} \dots P_{a_{L}}$ can be ordered using
the relation (\ref{pp1}) in order to achieve a common element. Thus we find
\begin{equation}
\label{ff2}
\sum_{\{ a_j \}} P_{a_{1}} P_{a_{2}} \dots P_{a_{L}} = (1+q^{-2})(1+q^{-2}+q^{-4}) \dots (1 + q^{-2} + q^{-4} + \dots + q^{-2(L-1)}) P_1 P_2 \dots P_L \; .
\end{equation}
Now in order to derive an explicit expression for the leading term coefficient in (\ref{a7}), we only
need to compute $\bra{\bar{0}} P_1 P_2 \dots P_L \ket{0}$. This task can be directly performed since $P_1 P_2 \dots P_L$ consists of a tensor
product of local operators. Using (\ref{vac}) and (\ref{a4}) we thus obtain
\begin{equation}
\label{ff3}
\bra{\bar{0}} P_1 P_2 \dots P_L \ket{0} = q^{\frac{L(L-1)}{2}} \; .
\end{equation}

Gathering our results so far, in particular the relations (\ref{ff1}), (\ref{ff2}) and (\ref{ff3}), we are left with
\begin{equation}
\bra{\bar{0}} \prod_{i=1}^{L} f_{L-1}^{(i)} \ket{0} = \frac{(q-q^{-1})^L}{2^{L^2}} \left[ L ! \right]_{q^2}
\end{equation}
where $\left[ L ! \right]_{q^2}$ denotes the $q$-factorial function defined as
\begin{equation}
[ L ]_{q^2} ! = 1 (1+q^2)(1+q^2+q^4) \dots (1 + q^2 + \dots + q^{2(L-1)}) \; .
\end{equation}

\addcontentsline{toc}{section}{Appendix B}
\section*{\bf Appendix B: The coefficients $h_{m,n,o}$}
\setcounter{equation}{0}
\renewcommand{\theequation}{B.\arabic{equation}}
In the table below we present the non-null coefficients $h_{m,n,o}$ of the polynomial (\ref{pol123}) solving the 
Eq. (\ref{L3}).
\begin{table}[!h]
\centering
\begin{tabular}{|c|c||c|c|}
\hline
$(m,n,o)$ & $\frac{h_{m,n,o}}{h_{2,2,2}}$ & $(m,n,o)$ & $\frac{h_{m,n,o}}{h_{2,2,2}}$ \\
\hline
$(-2,-2,-2)$ & $\frac{1}{q^6}$ & $(0, 0, 0)$ & $\frac{(1 - 8 q^2 - 34*q^4 - 8 q^6 + q^8)}{q^4 (1 + q^2) (1 - q + q^2) (1 + q + q^2)}$ \\
\hline 
$(-2, -2, 0)$ & $\frac{-3 (1 + q^2)}{q^4 (1 - q + q^2) (1 + q + q^2)}$ & $(0, 0, 2)$ & $\frac{12}{(1 - q + q^2)(1 + q + q^2)}$ \\
\hline 
$(-2, -2, 2)$ &  $\frac{3}{q^2 (1 - q + q^2)(1 + q + q^2)}$ & $(0, 2, -2)$ & $\frac{-(1 + 10 q^2 + q^4)}{q^2 (1 + q^2) (1 - q + q^2) (1 + q + q^2)}$ \\
\hline
$(-2, 0, -2)$ & $\frac{-3(1 + q^2)}{q^4 (1 - q + q^2)(1 + q + q^2)}$ & $(0, 2, 0)$ & $\frac{12}{(1 - q + q^2)(1 + q + q^2)}$ \\
\hline
$(-2, 0, 0)$ &  $\frac{12}{q^2 (1 - q + q^2)(1 + q + q^2)}$ & $(0, 2, 2)$ & $\frac{-3 (1 + q^2)}{(1 - q + q^2)(1 + q + q^2)}$ \\
\hline
$(-2, 0, 2)$ & $\frac{-(1 + 10 q^2 + q^4)}{q^2 (1 + q^2) (1 - q + q^2) (1 + q + q^2)}$ & $(2, -2, -2)$ & $\frac{3}{q^2 (1 - q + q^2) (1 + q + q^2)}$ \\
\hline
$(-2, 2, -2)$ & $\frac{3}{q^2 (1 - q + q^2) (1 + q + q^2)}$ & $(2, -2, 0)$ & $\frac{-(1 + 10 q^2 + q^4)}{q^2 (1 + q^2) (1 - q + q^2) (1 + q + q^2)}$ \\
\hline
$(-2, 2, 0)$ & $\frac{-(1 + 10 q^2 + q^4)}{q^2 (1 + q^2) (1 - q + q^2) (1 + q + q^2)}$ & $(2, -2, 2)$ & $\frac{3}{(1 - q + q^2) (1 + q + q^2)}$ \\
\hline
$(-2, 2, 2)$ & $\frac{3}{(1 - q + q^2) (1 + q + q^2)}$ & $(2, 0, -2)$ & $\frac{-(1 + 10 q^2 + q^4)}{q^2 (1 + q^2) (1 - q + q^2) (1 + q + q^2)}$ \\
\hline
$(0, -2, -2)$ & $\frac{-3 (1 + q^2)}{q^4 (1 - q + q^2) (1 + q + q^2)}$ & $(2, 0, 0)$ & $\frac{12}{(1 - q + q^2) (1 + q + q^2)}$ \\
\hline
$(0, -2, 0)$ & $\frac{12}{q^2 (1 - q + q^2)(1 + q + q^2)}$ & $(2, 0, 2)$ & $\frac{-3 (1 + q^2)}{(1 - q + q^2)(1 + q + q^2)}$ \\
\hline
$(0, -2, 2)$ & $\frac{-(1 + 10 q^2 + q^4)}{q^2 (1 + q^2) (1 - q + q^2) (1 + q + q^2)}$ & $(2, 2, -2)$ & $\frac{3}{(1 - q + q^2) (1 + q + q^2)}$ \\
\hline 
$(0, 0, -2)$ &  $\frac{12}{q^2 (1 - q + q^2)(1 + q + q^2)}$ & $(2, 2, 0)$ & $\frac{-3 (1 + q^2)}{(1 - q + q^2)(1 + q + q^2)}$ \\
\hline
\end{tabular}
\end{table}

\end{document}